\documentclass[aps,pra,twocolumn]{revtex4-2}

\usepackage{graphicx,amsmath,amssymb, color}

\DeclareMathOperator{\sgn}{sgn}

\begin{document}

\title{Optimal implementation of two-qubit linear optical quantum filters}

\author{Jarom\' ir Fiur\' a\v sek}
\affiliation{Department of Optics, Palack\' y University, 17. listopadu 1192/12,  771~46 Olomouc,  Czech Republic}

\author{Robert St\'{a}rek}
\affiliation{Department of Optics, Palack\' y University, 17. listopadu 1192/12,  771~46 Olomouc,  Czech Republic}

\author{Michal Mi\v{c}uda}
\affiliation{Department of Optics, Palack\' y University, 17. listopadu 1192/12,  771~46 Olomouc,  Czech Republic}

\begin{abstract}
We design optimal interferometric schemes for implementation of two-qubit linear optical quantum filters diagonal in the computational basis. 
The filtering is realized by interference of the two photons encoding the qubits in a multiport linear optical interferometer, followed by conditioning on presence of a single photon in each output port of the filter. 
The filter thus operates in the coincidence basis, similarly to many linear optical unitary quantum gates.  Implementation of the filter with linear optics may require an additional overhead in terms of reduced overall success probability 
of the filtering and the optimal filters are those that maximize the overall success probability. We discuss in detail the case of symmetric real filters and extend our analysis also to asymmetric and complex filters. 
\end{abstract}

\maketitle

\section{Introduction}

Quantum information processing with linear optics \cite{Kok2014,Slussarenko2019, Flamini2019} relies on encoding of qubits into states of single photons and implementation of various quantum operations by multiphoton interference,
followed by photon counting measurements and postselection based on the measurement outcomes. Scalable linear optical quantum gates can be in principle realized with the use of auxiliary single photons 
and feedforward operations controlled by the outcomes of measurements on auxiliary modes \cite{Knill2001,Kok2007}. During the past two decades, quantum information processing with linear optics has evolved rapidly, 
driven in recent years by important advances in design of integrated quantum optics circuits on a chip \cite{Wang2020},  highly efficient superconducting single-photon detectors \cite{Marsili2013,Jeannic2016} and single-photon sources \cite{Senellart2017,Meyer2020}. 
Although full-scale quantum computing with linear optics still appears to be technologically very demanding, the linear optics platform 
proved to be very useful for proof-of-principle tests of various concepts and protocols in quantum information processing, and small-scale linear optical quantum processors 
may find their applications in advanced quantum communication networks, where the role of light as the information carrier is indispensable.

A central topic in quantum computing with linear optics is to design and realize various two-qubit \cite{Kok2007} and multiqubit  \cite{Lanyon2009,Micuda2013,Patel2016,Ono2017,Starek2018} linear optical quantum gates.  
Besides unitary gates, non-unitary quantum operations, commonly referred to as quantum filters, also  play an essential role in quantum information processing. A quantum filter can be defined as a trace-decreasing completely 
positive map with a single Kraus operator $M$ that satisfies $M^\dagger M \leq I$ and transforms a general  input state $\rho_{\mathrm{in}}$ as $\rho_{\mathrm{out}}=M\rho_{\mathrm{in}}M^\dagger$. This output state is not normalized and 
 $P_S=\mathrm{Tr}[\rho_{\mathrm{out}}]$ is the probability of successful filtering. Quantum filters find their applications for instance in optimal quantum state discrimination \cite{Huttner1996,Clarke2001}, 
entanglement  concentration and distillation \cite{Bennett1996,Kwiat2001,Takahashi2010,Kurochkin2014}, 
 or in engineering of highly nonclassical states of light by conditional photon addition or subtraction \cite{Zavatta2004,Wenger2004,Ourjoumtsev2006,Kumar2013,Lvovsky2020}.

 In the present work we investigate optimal linear optical implementation of a two-qubit quantum filter diagonal in the computational basis,
 \begin{equation}
M=m_{00}|00\rangle\langle 00| +m_{01}|01\rangle\langle 01|+m_{10} |10\rangle\langle 10|+|11\rangle\langle 11|,
 \label{Kfilter}
 \end{equation}
 where $|m_{jk}|\leq 1$, and without loss of generality we set $m_{11}=1$. We concentrate on the resource-effective implementation that does not require any auxiliary photons. 
The filter is realized by interference of the two photons encoding the qubits in a suitably designed multiport optical interferometer,
 and successful filtering is heralded by presence of a single photon in each output of the filter. The filter thus operates in the coincidence basis, similarly to a number of linear optical unitary quantum gates designed and realized to date. In practice,
 the verification of presence of a single photon  in each output of the filter would require destructive coincidence two-photon detection. The quantum filters $M$ can be considered as generalization of two-qubit controlled-phase gates, where phase modulation is replaced by amplitude modulation. 
Specifically, for $m_{00}=1$ and $m_{01}=m_{10}=0$  the filter (\ref{Kfilter}) becomes the quantum parity check \cite{Pittman2001,Pittman2002,Hofmann2002} that is useful for implementation of the linear optical CNOT gate 
\cite{Pittman2001,Gasparoni2004,Zhao2005} and for generation of entangled multiphoton cluster states \cite{Browne2005}.

It turns out that, depending on the filter parameters, it may not be possible to implement the filter without additional reduction of probability of success. This means that instead of filter $M$ we implement an equivalent but less efficient filter $\sqrt{P_L}M$, where $P_L $ is the 
probability reduction factor imposed by the linear optical setup. Our goal is to design optimal interferometric schemes for the two-qubit quantum filters (\ref{Kfilter}), that maximize the probability $P_L$. This task is similar to the design of optimal two-qubit linear optical phase gates  
operating in the coincidence basis \cite{Kieling2010,Lemr2011}. However, in contrast to the optimal controlled-phase gate, we find that different mode-coupling configurations are optimal depending on the filter parameters. 
Importantly, fully analytical results can be obtained for the optimal interferometer parameters and the resulting maximum success probability $P_L$.

The rest of the paper is organized as follows. In Section II we present a general description of the linear optical interferometric scheme that implements  the two-qubit quantum filters. 
In Section III we discuss in detail realization of a symmetric filter with real coefficients and in Section IV we extend our analysis to asymmetric and complex filters.
Finally, Section V contains a brief summary and conclusions. The Appendix contains technical proof of the allowed structure of the interferometer that implements the quantum filter.

\section{Linear optical two-qubit quantum filters}

\begin{figure}[t!]
\includegraphics[width=0.8\linewidth]{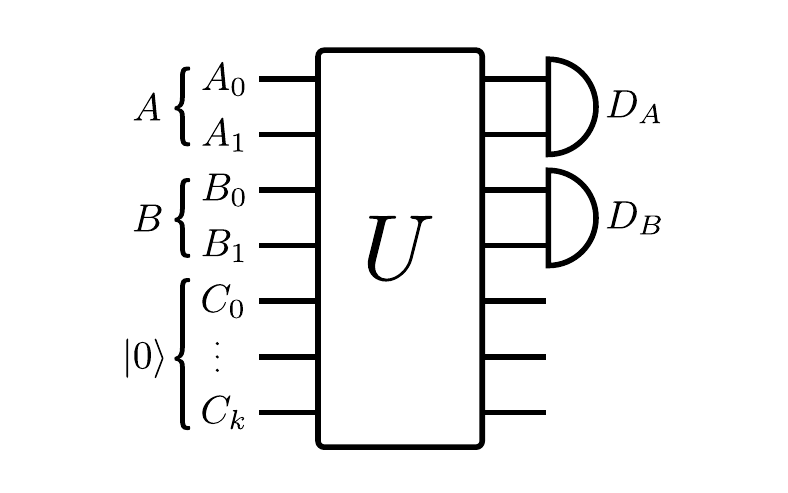}
\caption{Two-qubit linear optical quantum filter operating in the coincidence basis. Qubits are encoded into states of single photons in pairs of modes $A_0,A_1$ and $B_0,B_1$. 
The scheme involves also auxiliary input modes $C_j$ that are all prepared in a vacuum state. All modes are coupled in a multiport optical interferometer described by a unitary matrix $U$. 
Successful implementation of the filter is indicated by coincidence detection of two photons, one in modes $A_0,A_1$, the other in modes $B_0,B_1$.}
\label{fig-unitary}
\end{figure}

A conceptual scheme of linear optical setup implementing the two-qubit quantum filter (\ref{Kfilter}) is depicted in Fig.~\ref{fig-unitary}.
Each qubit is encoded into state of a single photon that can propagate in two modes denoted as $A_0$, $A_1$, and $B_0$, $B_1$ for the qubit A and B, respectively. Presence of a photon in mode $A_0$, $B_0$ 
represents logical state $|0\rangle$ while photon in mode $A_1$, $B_1$ encodes logical state $|1\rangle$.
The quantum filter is implemented by interference of the two photons in a multiport optical interferometer followed by verification of presence of a single photon in each pair of output modes $A_0,A_1$ and $B_0,B_1$.  
In practice, this verification can be performed destructively by conditioning on suitable two-photon coincidence detection. 
The linear optical quantum filter thus operates in the coincidence basis, similarly to certain linear optical two-qubit CNOT and controlled-phase gates \cite{Okamoto2005,Langford2005,Kiesel2005,Kok2007}.  

A multiport optical interferometer can be described by a unitary matrix $U$ that specifies the coupling between the input and output modes. Note that in addition to the four modes that encode the qubits the interferometer 
may contain also additional auxiliary modes.  In terms of creation operators $c_j^\dagger$ associated with each mode we have
\begin{equation}
c_{j,\mathrm{in}}^\dagger=\sum_{k} U_{j,k}c_{k,\mathrm{out}}^\dagger.
\end{equation}
Let $|1_{A_j},1_{B_k}\rangle$, where $j,k \in \{0,1\}$, denote the input two-photon Fock state corresponding to the two-qubit product state $|j\rangle_A |k\rangle_B$. 
Conditional on observation of a single photon in each pair of output modes $A_0$, $A_1$ and $B_0$, $B_1$,  the input state transforms according to 
\begin{equation}
|1_{A_j},1_{B_k}\rangle \rightarrow \sum_{m,n=0}^1 W_{A_m,B_n |A_j,B_k} |1_{A_m},1_{B_n}\rangle,
\end{equation}
where
\begin{equation}
 W_{A_m,B_n |A_j,B_k}=U_{A_j,A_m}U_{B_k,B_n}+U_{A_j,B_n}U_{B_k,A_m} .
\end{equation}
Correct implementation of the quantum filter (\ref{Kfilter}) requires that 
\begin{equation}
 W_{A_m,B_n |A_j,B_k} =\sqrt{P_L} m_{jk} \delta_{jm}\delta_{kn},
\label{Ucondition}
\end{equation}
where  $P_L \leq 1$ is an additional factor that may reduce the overall probability of implementation of the linear optical quantum filter. Our goal is to find for a given filter (\ref{Kfilter}) the optimal interferometer that maximizes $P_L$.

Design of the optimal  linear optical quantum filters is similar to the construction of optimal linear optical two-qubit quantum controlled-phase gate operating in the coincidence basis \cite{Kieling2010,Lemr2011}. 
In particular, one can show that only one pair of the information-encoding modes can be interferometrically coupled, 
see the Appendix for a proof. Consequently, the $4\times 4$  matrix  $U_{AB}=(U_{j,k})$, where $ j,k\in\{A_0,B_0,A_1,B_1\}$, has a block-diagonal structure, consisting of a general $2\times 2$ matrix describing coupling of two modes 
and two additional  diagonal elements specifying the amplitude transmittances for the other two modes. 
 For an explicit example of such matrix, see e.g. Eq. (\ref{UABmatrix00}) below. In what follows, we will frequently use the conditions under which a $2\times 2$ matrix $V=(V_{j,k})$ is a submatrix of a unitary matrix. Define two row vecors $v_j=(V_{j,0}, V_{j,1})$. Matrix $V$ 
is a submatrix of a unitary matrix if and only if  the vector norms and scalar product satisfy the  inequalities,
\begin{equation}
\begin{array}{c}
|\vec{v}_0|^2\leq 1, \qquad |\vec{v}_1|^2 \leq 1, \\[4mm]
|\vec{v}_0\cdot \vec{v}_1|^2 \leq (1-|\vec{v}_0|^2)(1-|\vec{v}_1|^2).
\label{vconditions}
\end{array}
\end{equation}
Here the last inequality guarantees that the vectors $\vec{v}_0$ and $\vec{v}_1$ can be completed to orthogonal vectors of unit length.

\section{Symmetric real filter}

In this section we investigate optimal interferometric schemes for implementation of a two-qubit real symmetric quantum filter specified by Kraus operator
\begin{equation}
M=a|00\rangle 00|+b(|01\rangle \langle 01|+|10\rangle \langle 10|)+|11\rangle\langle 11|,
\label{MfilterLO}
\end{equation}
where $0\leq a\leq 1$ and $0\leq b\leq 1$. Note that if $a=b^2$ then the filter factorizes and becomes a product of two single-qubit filters that each attenuate the amplitude of basis state $|0\rangle$ by factor $b$. 
Otherwise,  the quantum filter is an entangling operation that can create entangled states from input separable states. When optimizing the success probability $P_L$, it is necessary 
to consider three different configurations: coupling of modes $A_0$ and $B_0$, coupling of modes $A_1$ and $B_1$, and finally also coupling of modes $B_0$ and $A_1$.
 Note that due to the symmetry of the filter, the fourth possible configuration, where modes $B_1$ and $A_0$ are coupled, is fully equivalent to the configuration where modes $A_1$ and $B_0$
are coupled, and therefore need not be considered separately.  In what follows we discuss each of the above listed configurations in detail.

\subsection{Coupling of modes $A_0$ and $B_0$}
\label{symmetricA0B0}

Assuming coupling of modes $A_0$ and $B_0$ and the ordering of modes $A_0$, $B_0$, $A_1$, $B_1$ we can write the corresponding $4\times 4$ submatrix of $U$ as follows (see the Appendix),
\begin{equation}
U_{AB}=\left(
\begin{array} {cccc}
\tau_A b & \tau_A x & 0 & 0 \\
\tau_B y & \tau_B b & 0 & 0 \\
0  & 0 & \tau_A & 0 \\
0  & 0 & 0 & \tau_B 
\end{array}
\right).
\label{UABmatrix00}
\end{equation}
Here $\tau_A$ and $\tau_B$ represent the amplitude attenuation of modes $A_1$ and $B_1$, respectively, and the parameters $x$ and $y$ specify the strength of the interferometric coupling between modes $A_0$ and $B_0$.
Without loss of generality, we can assume that all matrix elements of $U_{AB}$ are real. The parameters $x$ and $y$ are related by the condition
\begin{equation}
xy=a-b^2.
\label{xy00}
\end{equation}
We are thus left with three free parameters  $\tau_A$, $\tau_{B}$ and $x$ that shall be optimized to maximize the probability
\begin{equation}
P_L=\tau_A^2\tau_B^2.
\end{equation} 
Since $U_{AB}$ is a submatrix of a unitary matrix, the following constraints must be satisfied (c.f. also Eq. (\ref{vconditions})),
\begin{equation}
\tau_A^2\leq 1 \qquad \tau_B^2\leq 1.
\label{tauABuniversal}
\end{equation}
\begin{equation}
\tau_A^2(b^2+x^2)\leq 1, \quad \tau_B^2(b^2+y^2)\leq 1, 
\label{tauAB00}
\end{equation}
and
\begin{equation}
\tau_A^2\tau_B^2b^2(x+y)^2 \leq[1-\tau_A^2(b^2+x^2)][1-\tau_B^2(b^2+y^2)].
\label{scalar00}
\end{equation}
Taking into account the constraint (\ref{xy00}), and introducing new parameters $z$, $\tau_B=z\tau_A$, and  $\gamma=|y/x|z $, this last inequality can be rewritten as 
\begin{equation}
 \sqrt{P_L}b^2(z^{-1}+z)+\sqrt{P_L}|a-b^2|(\gamma^{-1}+\gamma)-P_L(2b^2-a)^2 \leq 1.
\end{equation}
 Since 
\begin{equation}
x+x^{-1}\geq 2, \quad \forall x>0,
\label{meansinequality}
\end{equation}
 we get
\begin{equation}
2\sqrt{P_L}(b^2+|a-b^2|)-P_L (2b^2-a)^2 \leq 1.
\label{scalarbound00}
\end{equation}
This inequality yields a nontrivial upper bound on $P_L$ if $a>b^2$. Assuming equality in Eq. (\ref{scalarbound00}), and carefully analyzing the two roots of the resulting quadratic equation for $\sqrt{P_L}$,
\begin{equation}
\sqrt{P_L}= \frac{a\pm 2b\sqrt{a-b^2}}{(2b^2-a)^2}=\frac{1}{(b \mp \sqrt{a-b^2})^2},
\end{equation}
we find that $P_L$ is upper bounded by the smaller root, and
\begin{equation}
P_{L}\leq \frac{1}{(b+\sqrt{a-b^2})^4} , \qquad a>b^2.
\label{scalarbound00final}
\end{equation}
Another useful inequality can be obtained by taking the product of the two inequalities (\ref{tauAB00}). We get
\begin{equation}
P_L[b^4+b^2|b^2-a|(\mu+\mu^{-1})+(b^2-a)^2] \leq 1,
\end{equation}
where $\mu=|x/y|$. With the use of inequality (\ref{meansinequality}) this yields
\begin{equation}
P_L \leq \frac{1}{\left(b^2+|b^2-a|\right)^{2}}.
\label{PLboundtau00}
\end{equation}
We now explicitly present the optimal interferometric configurations that are all symmetric, $\tau_A=\tau_B=\tau$ and $x=\pm y$. We have to distinguish four different cases according to the values of the filter parameters $a$ and $b$.

\begin{figure}[!t!]
\includegraphics[width=\linewidth]{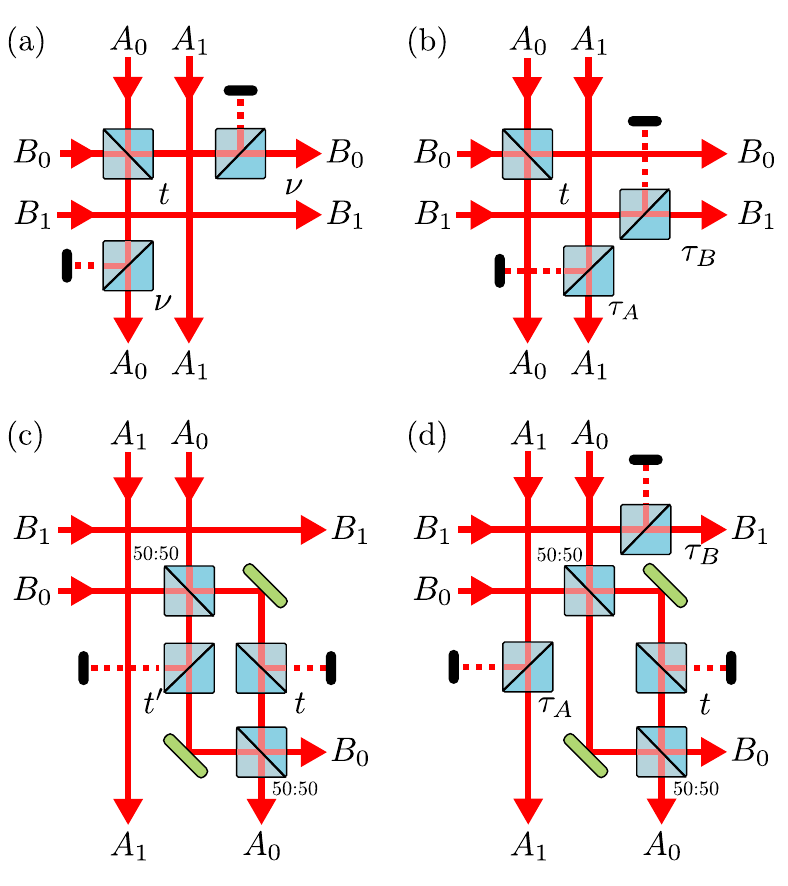}
\caption{Optimal optical interferometers implementing the two-qubit quantum filters (\ref{MfilterLO})  by interferometric coupling of modes $A_0$ and $B_0$. The labels of beam splitters indicate their amplitude transmittances. 
Mode attenuation is realized by propagation through a beam splitter with suitable transmittance, whose auxiliary input mode is prepared in vacuum state. The four schemes (a)-(d) represent optimal setups for different values of filter 
parameters $a$  and $b$. For details, see text.}
\label{fig-LOfilters}
\end{figure}

(i) $a\leq b^2$, $2b^2-a<1$. As shown in Fig.~\ref{fig-LOfilters}(a), in this case it is optimal to couple the modes $A_0$ and $B_0$ on a beam splitter with amplitude transmittance
\begin{equation}
t=\frac{b}{\sqrt{2b^2-a}}.
\end{equation}
Subsequently, each of the modes $A_0$ and $B_0$ is attenuated with amplitude factor
\begin{equation}
\nu=\sqrt{2b^2-a}
\end{equation}
by sending it through a beam splitter with amplitude transmittance $\nu$ whose auxiliary mode is prepared in vacuum state. 
In this case $P_L=1$ and the linear optical implementation does not impose any extra reduction of the overall success probability of the quantum filtering.

(ii) $a\leq b^2$, $2b^2-a>1$. The optimal scheme is drawn in Fig.~\ref{fig-LOfilters}(b) and is similar to that in case (i). However, instead of attenuating modes $A_0$ and $B_0$ we have to attenuate modes $A_1$ and $B_1$ with amplitude transmittance 
\begin{equation}
\tau_A=\tau_B=\frac{1}{\sqrt{2b^2-a}}.
\end{equation}
Subsequently, the probability $P_L$ drops below $1$ and we get $P_L=(2b^2-a)^{-2}$. The scheme is optimal because $P_L$ saturates the inequality (\ref{PLboundtau00}).

(iii) $a> b^2$, $b+\sqrt{a-b^2}\leq 1$. The optimal interferometric scheme is shown in Fig.~\ref{fig-LOfilters}(c). Modes $A_0$ and $B_0$ are injected into a Mach-Zehnder interferometer formed by two balanced beam splitters. 
One arm of the interferometer is attenuated with amplitude transmittance $t$ and the other with amplitude transmittance $t'$, where 
\begin{equation}
t=b-\sqrt{a-b^2},
\qquad
t'=b+\sqrt{a-b^2}.
\end{equation}
In this case we achieve $P_L=1$.

(iv)  $a> b^2$, $b+\sqrt{a-b^2} > 1$. The optimal scheme is shown in Fig.~\ref{fig-LOfilters}(d) and is similar to the scheme for case (iii). However, only one of the interferometer arms is attenuated, with amplitude transmittance
\begin{equation}
 t= \frac{b-\sqrt{a-b^2}}{b+\sqrt{a-b^2}}.
\end{equation} 
Furthermore, modes $A_1$ and $B_1$ are each attenuated by factor
\begin{equation}
\tau_A=\tau_B= \frac{1}{b+\sqrt{a-b^2}}.
\end{equation}
Consequently, we have
\begin{equation}
P_L= \frac{1}{(a+2b\sqrt{a-b^2})^2}.
\end{equation}
The scheme is optimal because the achieved $P_L$ saturates the bound (\ref{scalarbound00final}).

\subsection{Coupling of modes $A_1$ and $B_1$}
Let us now investigate the configuration where modes $A_1$ and $B_1$ are interferometrically coupled instead of the modes $A_0$ an $B_0$. Keeping the same ordering of modes
 $A_0$, $B_0$, $A_1$, $B_1$,  the relevant $4 \times 4$ submatrix of $U$ can be written as
\begin{equation}
U_{AB}=\left(
\begin{array} {cccc}
 \tau_A & 0 & 0  & 0 \\
0 & \tau_B & 0  & 0  \\
0 & 0 &  \frac{b}{a}\tau_A  & \tau_A x  \\
0 & 0  & \tau_B y & \frac{b}{a}\tau_B  \\
\end{array}
\right),
\label{UAB11}
\end{equation}
where 
\begin{equation}
xy=\frac{a-b^2}{a^2}
\label{xy11}
\end{equation}
and the probability $P_L$  can be expressed as
\begin{equation}
P_L=\frac{\tau_A^2\tau_B^2}{a^2}.
\end{equation}
The conditions following from the requirement that (\ref{UAB11}) is a submatrix of a unitary matrix yield
\begin{eqnarray}
\tau_A^2\left( \frac{b^2}{a^2}+x^2  \right)\leq 1, \quad 
\tau_B^2\left( \frac{b^2}{a^2}+y^2  \right)\leq 1, 
\label{tauinequalities11}
\end{eqnarray}
and
\begin{equation}
2\frac{\sqrt{P_L}}{a}(b^2+|a-b^2|)-\frac{P_L}{a^2}(2b^2-a)^2 \leq 1.
\label{scalarbound11}
\end{equation}
This last inequality was obtained by the same procedure as the inequality (\ref{scalarbound00}) and it implies the following upper bound on $P_L$,
\begin{equation}
P_L \leq \frac{a^2}{(b+\sqrt{a-b^2})^4},\qquad a>b^2.
\label{scalarboundfinal11}
\end{equation}
By taking the product of the two inequalities (\ref{tauinequalities11}) and utilizing the constraint (\ref{xy11}) we find that
\begin{equation}
P_L\leq \frac{a^2}{(b^2+|b^2-a|)^2}.
\label{productbound11}
\end{equation}
For $a<b^2$ this bound is stricter than the bound (\ref{PLboundtau00}). Similarly, for $a>b^2$ the inequality (\ref{scalarboundfinal11}) is stricter than the inequality (\ref{scalarbound00final}).
It follows from the inequalities (\ref{scalarboundfinal11}) and (\ref{productbound11}) that with the coupling of modes $A_1$ and $B_1$ we can achieve $P_L=1$ only if $a=b^2$. Physically, for $a\neq b^2$  there will always be a nonzero probability that for the input state $|11\rangle$ 
the two photons will bunch and will both end up either in mode $A_1$ or $B_1$, resulting in the failure of the filter.  We can therefore conclude that the interferometric coupling of modes $A_1$ and $B_1$ cannot yield higher $P_L$ 
than coupling of modes $A_0$ and $B_0$.

\subsection{Coupling of modes $B_0$ and $A_1$}
Finally, we consider an asymmetric configuration where modes $B_0$ and $A_1$ are coupled. The corresponding matrix $U_{AB}$ can be expressed as
\begin{equation}
U_{AB}=\left(
\begin{array} {cccc}
b\tau_A  & 0 & 0  & 0 \\
0 & \tau_B \frac{a}{b} & \tau_B x  & 0 \\
0 &\tau_A y & \tau_A  & 0 \\
0  & 0 & 0 & \tau_B 
\end{array}
\right),
\label{UABmatrix01}
\end{equation}
where
\begin{equation}
xy=b-\frac{a}{b}.
\label{xy01}
\end{equation}
The requirement that $U_{AB}$ is a submatrix of a unitary matrix yields the constraints
\begin{equation}
\tau_A^2(1+y^2)\leq 1, \qquad 
\tau_B^2\left(x^2 +\frac{a^2}{b^2}\right)\leq 1,
\label{tauAB01}
\end{equation}
and
\begin{equation}
\tau_A^2\tau_B^2\left(x+y\frac{a}{b}\right)^2 \leq \left[1-\tau_A^2(1+y^2)\right]\left[ 1-\tau_B^2\left(x^2+\frac{a^2}{b^2}\right) \right],
\label{scalar01}
\end{equation}
together with $\tau_A^2\leq 1$ and $\tau_B^2\leq 1$ .
The optimal schemes must saturate at least one of the inequalities (\ref{tauAB01}) and (\ref{scalar01}). If none of these inequalities is saturated, then we can increase $\tau_A$, hence also $P_L=\tau_A^2\tau_B^2$, until at least one inequality is saturated.

\begin{figure}
\includegraphics[width=0.9\linewidth]{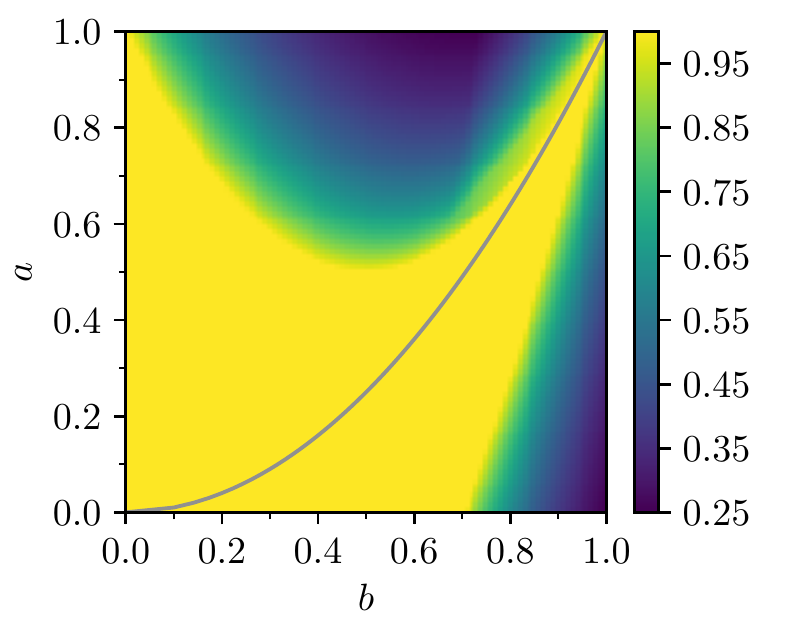}
\caption{Maximum probability $P_L$ for the symmetric two-qubit quantum filters (\ref{MfilterLO}) achievable with linear optical interferometric schemes is plotted as a function of filter parameters $a$ and $b$. The large yellow area represents filters for which $P_L=1$. 
The gray line indicates the points $a=b^2$ that correspond to filters which factor into products of single-qubit filters.}
\label{fig-PL}
\end{figure}

Let us first assume that one of the inequalities (\ref{tauAB01}) is saturated. It immediately follows from Eq. (\ref{scalar01}) that $x+ya/b=0$ must hold, which together with (\ref{xy01}) yields
\begin{equation}
x=\frac{\sqrt{a}}{b}\sqrt{a-b^2}, \qquad y=-\sqrt{\frac{a-b^2}{a}}.
\label{xyoptimalB0A1branch1}
\end{equation}
This solution exists in the parameter  region $a>b^2$. It follows from the inequalities (\ref{tauAB01}) that the maximum possible values of $\tau_{A,B}$ are given by
\begin{equation}
\tau_A^2=\frac{a}{2a-b^2}, \qquad \tau_B^2 =\min\left(1,\frac{b^2}{a(2a-b^2)}\right).
\end{equation}
Consequently, the maximum achievable $P_L$ for this case can be expressed as
\begin{equation}
P_L=\min \left(\frac{a}{2a-b^2},\frac{b^2}{(2a-b^2)^2}\right), \quad a>b^2.
\end{equation}
Let us now assume that only the inequality (\ref{scalar01}) is saturated. Since the saturation means that equality holds in (\ref{scalar01}), we can use it to express $\tau_A^2$  in terms of $x$ and $\tau_B$,
\begin{equation}
\tau_A^2=\frac{x^2[b^2-\tau_B^2(b^2x^2+a^2)]}{x^2b^2+(b^2-a)^2-x^2\tau_B^2(2a-b^2)^2}.
\label{tauA2A1B0}
\end{equation}
The optimal values of $\tau_B$ and $x$ can be determined by solving the extremal equations 
\begin{equation}
\frac{\partial}{\partial \tau_B}  (\tau_A^2\tau_B^2)=0, \qquad \frac{\partial}{\partial x}( \tau_A^2\tau_B^2)=0,
\end{equation}
where $\tau_A^2$ is given by Eq. (\ref{tauA2A1B0}). In the region $a>b^2$ we recover the optimality condition (\ref{xyoptimalB0A1branch1}). In the region $a<b^2$ we obtain additional potentially optimal solution
\begin{equation}
x=\frac{\sqrt{a}}{b}\sqrt{b^2-a}, \qquad y=\sqrt{\frac{b^2-a}{a}},
\end{equation}
and
\begin{equation}
\tau_A^2=\frac{a}{(\sqrt{b^2-a}+\sqrt{a})^2},   \qquad \tau_B^2=\frac{b^2}{a}\frac{1}{(\sqrt{b^2-a}+\sqrt{a})^2}.
\end{equation}
Note that this solution is acceptable only if all the inequalities (\ref{tauABuniversal}) and (\ref{tauAB01}) are satisfied. Additionally, we have to consider also the extremal point $\tau_B^2=1$. On inserting this into Eq. (\ref{tauA2A1B0}), we have 
\begin{equation}
P_L=\tau_A^2=\frac{x^2(b^2-b^2x^2-a^2)]}{x^2 b^2+(b^2-a)^2-x^2(2a-b^2)^2}.
\end{equation}
The optimal $x$ maximizing $P_L$ can be found from the extremal equation
\begin{equation}
\frac{\partial P_L}{\partial x}=0.
\end{equation}
After some algebra, this yields two roots
\begin{equation}
x^2=\frac{(a+b)(a-b^2)}{b(2a+b-b^2)}, \qquad P_L=\frac{(a+b)^2}{(2a+b-b^2)^2},
\label{solution1}
\end{equation}
and
\begin{equation}
x^2=\frac{(b-a)(b^2-a)}{b(b^2+b-2a)}, \qquad P_L=\frac{(a-b)^2}{(b^2+b-2a)^2}.
\label{solution2}
\end{equation}
We emphasize that the formulas (\ref{solution1}) or (\ref{solution2}) represent valid potential optimal points only if $x^2\geq 0$ and if all the inequalities  (\ref{tauABuniversal}) and (\ref{tauAB01})  are satisfied.

\begin{figure}
\centerline{\includegraphics[width=0.8\linewidth]{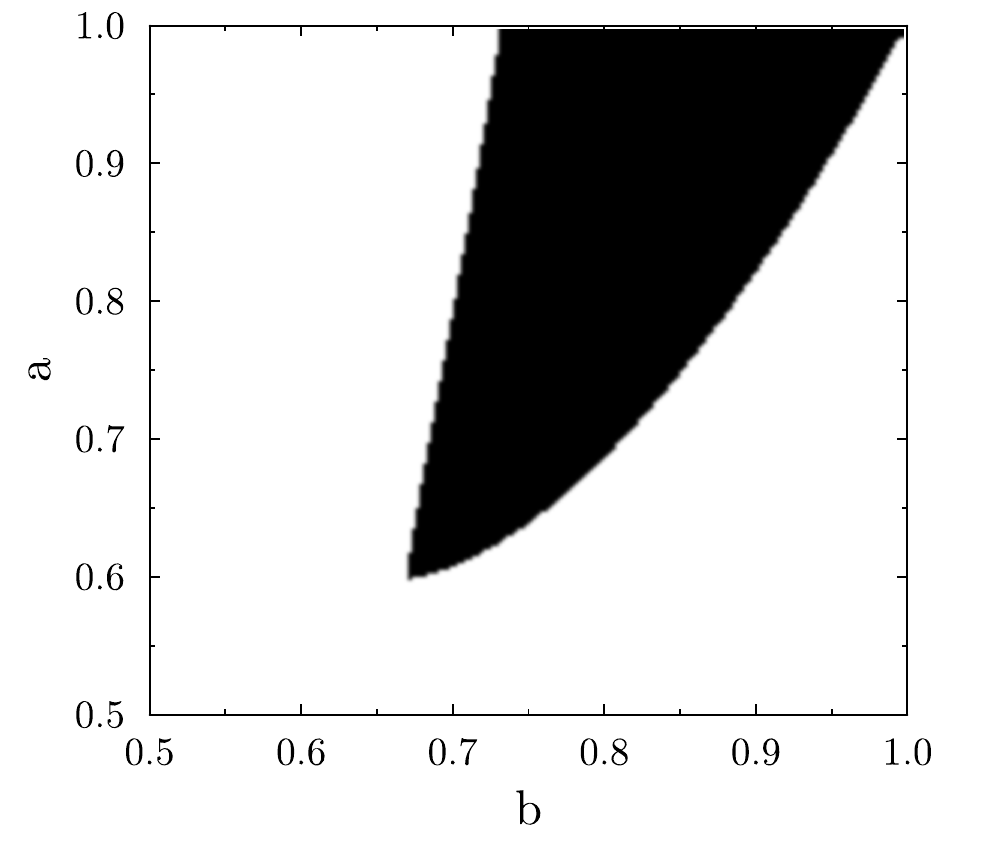}}
\caption{The black area indicates the range of parameters $a$ and $b$ of a symmetric two-qubit quantum filter (\ref{MfilterLO}) for which the coupling of modes $B_0$ and  $A_1$ leads to maximum success probability $P_L$.}
\label{fig-branches}
\end{figure}

The final maximization of $P_L$ must be performed over all the above considered configurations and all the identified potentially optimal solutions. 
The maximal $P_L$, optimized over all the coupling configurations, is plotted in Fig.~\ref{fig-PL}. Remarkably, we find that for a certain range of parameters $a$ and $b$ satisfying $a>b^2$ the asymmetric  scheme where modes  $B_0$ and $A_1$  are coupled outperforms the symmetric scheme 
where modes $A_0$ and $B_0$ are coupled, and achieves higher $P_L$. This area of parameters where the coupling of modes $B_0$ and $A_1$ is optimal is depicted in Fig.~\ref{fig-branches}.
 We note that the interferometric coupling described by matrix (\ref{UABmatrix01}) can be realized by interference of modes $B_0$ and $A_1$ in a Mach-Zehneder interferometer  
formed by two generally unbalanced beam splitters, and the signal in each interferometer arm should be suitably attenuated, c.f. Fig.~\ref{fig-LOfilters}(c).
The splitting ratios of the beam splitters and the attenuation factors can be determined by singular value decomposition of the matrix $U_{AB}$ \cite{Lemr2011}.

\section{Asymmetric and complex filters}
The optimization procedure discussed in the previous section can be extended to asymmetric and complex two-qubit filters. Here we illustrate it on the examples of two-qubit asymmetric filter with real coefficients and a two-qubit symmetric complex filter. 
We shall  focus on the configuration where modes $A_0$ and $B_0$ are coupled. Configurations where other pairs of modes are coupled can be treated in a similar manner. For an asymmetric filter one has to consider separately both coupling of modes $A_0,B_1$ and $A_1,B_0$ because the symmetry is broken. 

\subsection{Asymmetric real filter}
Let us consider linear optical implementation of an asymmetric real filter 
\begin{equation}
M=a|00\rangle\langle 00| +b_A|01\rangle\langle 01|+b_B|10\rangle \langle 10|+|11\rangle\langle 11|,
\label{Masym}
\end{equation}
where $0\leq a \leq 1$, and  $0\leq b_A\leq b_B \leq 1$ are real parameters of the filter.  Assuming coupling of modes $A_0$ and $B_0$,  the matrix $U_{AB}$ can be conveniently parameterized as
\begin{equation}
U_{AB}=
\left(
\begin{array}{cccc}
b_A \tau_A & b_A\tau_A x & 0 & 0 \\
b_B \tau_B y & b_B \tau_B & 0 & 0 \\
0 & 0 & \tau_A & 0 \\
0 & 0 & 0 & \tau_B \\
\end{array}
\right),
\end{equation}
where
\begin{equation}
xy=\frac{a}{b^2}-1,
\label{xyasym}
\end{equation}
and we have defined the parameter $b=\sqrt{b_A b_B}$. Since $U_{AB}$  is a submatrix of a unitary matrix, the following inequalities must hold, similarly to the previously studied case of symmetric filter:
\begin{equation}
\tau_A^2\leq 1, \qquad \tau_{B}^2\leq 1,
\label{asymtauAB1}
\end{equation}
\begin{equation}
b_A^2\tau_A^2(1+x^2)\leq 1 , \qquad b_B^2\tau_B^2(1+y^2)\leq 1 , 
\label{asymtauAB2}
\end{equation}
and
\begin{equation}
\eta_A^2\eta_B^2(x+y)^2\leq [1-\eta_A^2(1+x^2)][1-\eta_B^2(1+y^2)],
\label{asymscalarprimary}
\end{equation}
where $\eta_A=b_A\tau_A$ and $\eta_B=b_B\tau_B$. With the use of condition (\ref{xyasym}), the last inequality can be rewritten as
\begin{equation}
\eta_A^2(1+x^2)+\eta_B^2(1+y^2)-\eta_A^2\eta_B^2\left(2-\frac{a}{b^2}\right)^2 \leq 1.
\label{asymscalar}
\end{equation}
For any filter (\ref{Masym}), the optimal interferometer maximizing $P_L=\tau_A^2\tau_B^2$ can always be designed such that the inequality (\ref{asymscalar}) is saturated and equality holds. Fist note that if one of the inequalities (\ref{asymtauAB2}) is saturated, then also inequality (\ref{asymscalarprimary}) 
is saturated and equality must hold, because  both the left and righ-hand sides of Eq. (\ref{asymscalarprimary}) must be equal to $0$. Assume now a configuration where none of the inequalities (\ref{asymtauAB2}) and (\ref{asymscalarprimary}) is saturated. 
If $\tau_A$ or $\tau_B$ is smaller than $1$, then we can increase their value until either equality  holds in (\ref{asymscalarprimary}) or $\tau_A=\tau_B=1$. For an optimal configuration with $\tau_A=\tau_B=1$ we can increase or decrease the free parameter $x$ while keeping the constraint (\ref{xyasym}) 
until equality holds in Eq. (\ref{asymscalarprimary}).

We now discuss the various options that have to be considered. Let us first consider the option $\tau_A=\tau_B=1$, i.e. $P_L=1$. In this case, $x$ and $y$ can be determined by solving Eqs. (\ref{xyasym}) and (\ref{asymscalar}), where equality is assumed to hold. 
We obtain
\begin{eqnarray}
x^2&=&\frac{1}{2b_A^2}\left[q+\sqrt{q^2-4(a-b^2)^2}\right], \nonumber \\
y^2&=&\frac{1}{2b_B^2}\left[q-\sqrt{q^2-4(a-b^2)^2}\right],
\label{asymxycase1}
\end{eqnarray}
where 
\begin{equation}
q=1-b_A^2-b_B^2+(2b^2-a)^2.
\end{equation}
Since $x^2$ and $y^2$ must be real and non-negative, the solution (\ref{asymxycase1}) exists only if
\begin{equation}
q \geq 2|a-b^2|.
\label{asymqinequality}
\end{equation}
Additionally, the inequalities (\ref{asymtauAB2}) must also hold, which reduces to 
\begin{equation}
x^2\leq \frac{1}{b_A^2}-1, \qquad  y^2\leq \frac{1}{b_B^2}-1,
\label{asymxyinequalitycase1}
\end{equation}
where $x^2$ and $y^2$ are given by Eq. (\ref{asymxycase1}). To sum up, $P_L=1$ is achievable with coupling of modes $A_0$ and $B_0$ if and only if the inequalities (\ref{asymqinequality}) and (\ref{asymxyinequalitycase1}) are satisfied.

Let us now assume that $\tau_A=1$ but $\tau_B$ can be smaller than $1$. Assuming equality in Eq. (\ref{asymscalar}) we get,
\begin{equation}
\tau_B^2=\frac{b_A^2 x^2[1-b_A^2(1+x^2)]}{b^4x^2+(a- b^2)^2-b_A^2x^2\left(2b^2-a\right)^2}.
\label{asymtauB2case2}
\end{equation}
The optimal $x^2$ that maximizes $\tau_B^2$ can be determined by solving the extremal equation
\begin{equation}
\frac{\partial \tau_B^2}{\partial x}=0.
\end{equation}
This leads to quadratic equation for $x^2$ with roots 
\begin{eqnarray}
x^2&=&\frac{(a-b^2)(1-b_A)}{b_A(b^2+ab_A-2b^2b_A)}, \nonumber \\
x^2&=&\frac{(b^2-a)(1+b_A)}{b_A(b^2-ab_A+2b^2b_A)}.
\end{eqnarray}
These roots 
represent valid solutions provided that $x^2 \geq 0$ and the inequalities (\ref{asymtauAB1}) and (\ref{asymtauAB2}) are satisfied, where $y^2$ and $\tau_B^2$ are determined by Eqs. (\ref{xyasym}) and (\ref{asymtauB2case2}), respectively.
For an asymmetric filter we must also independently consider the configuration $\tau_B=1$ because the symmetry is broken. Following a similar procedure as before, we obtain
\begin{equation}
\tau_A^2=\frac{b_B^2y^2[1-b_B^2(1+y^2)]}{b^4y^2+(a- b^2)^2-b_B^2y^2\left(2b^2-a\right)^2},
\label{asymtauA2case2}
\end{equation}
and the potentially optimal $y^2$ read
\begin{eqnarray}
y^2&=&\frac{(a-b^2)(1-b_B)}{b_B(b^2+ab_B-2b^2b_B)}, \nonumber \\
y^2&=&\frac{(b^2-a)(1+b_B)}{b_B(b^2-ab_B+2b^2b_B)}.
\end{eqnarray}
Once again these solutions are valid only if $y^2 \geq 0$ and the inequalities (\ref{asymtauAB1}) and (\ref{asymtauAB2}) are satisfied.

Finally, we consider the configuration where both $\tau_A$ and $\tau_B$ can be smaller than $1$. Assuming equality in (\ref{asymscalar}), we can express $\tau_A$ as a function of $\tau_B$ and $x$,
\begin{equation}
\tau_A^2=\frac{b_B^2 y^2[1-b_B^2\tau_B^2(1+y^2)]}{b^4y^2+(a-b^2)^2-b_B^2\tau_B^2 y^2(2b^2-a)^2}.
\end{equation}
On inserting this formula into the extremal equations 
\begin{equation}
\frac{\partial }{\partial \tau_B}(\tau_A^2\tau_B^2) =0, \qquad \frac{\partial }{\partial y}(\tau_A^2\tau_B^2) =0,
\label{asymextremal}
\end{equation}
we obtain after some algebra the following expressions for $x$ and $y$,
\begin{equation}
x=\frac{\sqrt{|a-b^2|}}{b}, \quad y=\sgn(a-b^2)\frac{\sqrt{|a-b^2|}}{b}.
\end{equation}
If $a \leq b^2$, then $x=-y$ and at least one of the inequalities (\ref{asymtauAB2}) is saturated. However, the inequalities (\ref{asymtauAB1}) may represent an additional bound. We can succinctly express  $\tau_A$ and $\tau_B$ as follows,
\begin{eqnarray}
\tau_A^2&=& \min\left(1,\frac{b^2}{b_A^2(2b^2-a)}\right), \nonumber \\
 \tau_B^2&=&\min\left(1,\frac{b^2}{b_B^2(2b^2-a)}\right).
\end{eqnarray}
If $a>b^2$, then the extremal equations (\ref{asymextremal}) lead to the following expressions for $\tau_A$ and $\tau_B$,
\begin{equation}
\tau_A^2=\frac{b_B}{b_A}\frac{1}{(\sqrt{a-b^2}+b)^2}, \qquad  \tau_B^2=\frac{b_A}{b_B}\frac{1}{(\sqrt{a-b^2}+b)^2}. 
\end{equation}
These formulas represent valid solutions only if the inequalities (\ref{asymtauAB1}) and (\ref{asymtauAB2}) are satisfied.

\begin{figure*}
	\centerline{\includegraphics[width=0.95\linewidth]{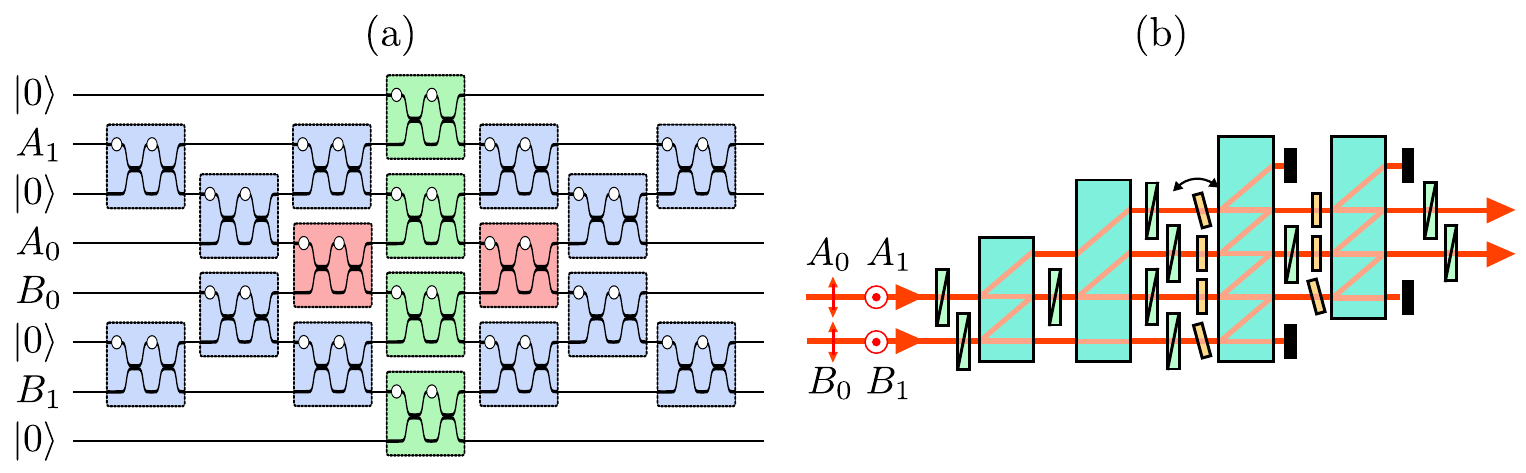}}
	\caption{Two examples of a possible implementation of the two-qubit linear optical quantum filters. (a) On-chip implementation with path encoding of the qubit states. Lines represent the optical waveguides and their crossings balanced directional couplers. 
The empty circles indicate variable phase shifters.  Sub-blocks in colored boxes act as variable beam splitters.  Blue boxes serve for mode swapping that ensures coupling of the desired pair of modes. 
The two red boxes realize the required interferometric coupling of the selected pair of signal carrying modes, and the four green boxes serve for tunable signal attenuation in each mode. 
(b) Bulk-optics implementation with polarization encoding and interferometers formed by calcite beam displacers that introduce lateral shift between the vertically and horizontally polarized beams. 
The polarization states are transformed with half-wave plates (green elements) that play the role of beam splitters, and the interferometric phase shifts can be set and controlled by tilting thin glass plates (orange elements). }
	\label{fig-integrated}
\end{figure*}

\subsection{Symmetric complex filter}
Let us finally investigate realization of symmetric two-qubit filters with complex coefficients. Without loss of generality, we can restrict ourselves to the filters
\begin{equation}
M=a e^{i\varphi} |00\rangle \langle 00| +b|01\rangle \langle 01|+b|10\rangle \langle 10|+|11\rangle \langle 11|,
\end{equation}
where $a$ and $b$ are real and positive, because the relative phase shifts of states $|01\rangle$ and $|10\rangle$ can be set to zero by suitable phase shifts applied to modes $A_0$ and $B_0$, respectively.
We shall again focus on the configuration where modes $A_0$ and $B_0$ are coupled. The matrix $U_{AB}$ has the same structure as for real symmetric filters, 
\begin{equation}
U_{AB}=
\left(
\begin{array}{cccc}
b \tau_A &  \tau_A x & 0 & 0 \\
 \tau_B y & b \tau_B & 0 & 0 \\
0 & 0 & \tau_A & 0 \\
0 & 0 & 0 & \tau_B \\
\end{array}
\right),
\end{equation}
only the condition on parameters $x$ and $y$ changes to,
\begin{equation}
xy=a e^{i\varphi}-b^2.
\end{equation}
Since $x$ and $y$ are generally complex, the conditions implied by $U_{AB}$ being a submatrix of a unitary matrix must be written as follows,
\begin{equation}
\tau_A^2(b^2+|x|^2)\leq 1, \qquad b^2\tau_A^2(b^2+|y|^2)\leq 1, 
\label{complexnorms}
\end{equation}
and
\begin{equation}
b^2 \tau_A^2\tau_B^2 |x+y^\ast|^2 \leq [1-\tau_A^2(b^2+|x|^2)][1-\tau_B^2(b^2+|y|^2)].
\label{complexscalar}
\end{equation}
Taking into account the symmetry of the filter, one can show that $P_L=1$ can be achieved provided that the inequalities (\ref{complexnorms}) and (\ref{complexscalar}) are satisfied for a symmetric configuration with $|x|=|y|$ and $\tau_A=\tau_B=1$. After some algebra, this yields the following condition,
\begin{equation}
b^2+b\sqrt{2(s+a\cos\varphi-b^2)}+s\leq 1.
\label{complexqinequality}
\end{equation}
where 
\begin{equation}
s=|xy|=\sqrt{a^2+b^4-2ab^2\cos\varphi}.
\end{equation}
If the inequality (\ref{complexqinequality}) does not hold, then the optimal configuration is symmetric, with
\begin{equation}
x=y=\sqrt{ae^{i\varphi}-b^2},
\end{equation}
and
\begin{equation}
\tau_A^2=\tau_B^2=\frac{1}{b^2+b\sqrt{2(s+a\cos\varphi-b^2)}+s}.
\end{equation}
This yields 
\begin{equation}
P_L=\left[b^2+b\sqrt{2(s+a\cos\varphi-b^2)}+s\right]^{-2}.
\label{complexPL}
\end{equation}
For $\varphi=0$ we recover the results for symmetric real filter derived in Section \ref{symmetricA0B0}. Also, for $a=b=1$ we recover from Eq. (\ref{complexPL}) 
the maximum probability of implementation of a two-qubit linear optical controlled-phase gate \cite{Kieling2010,Lemr2011},
\begin{equation}
P_{\mathrm{CP}}(\varphi)=\left[1+2\left|\sin\frac{\varphi}{2}\right| +2\sqrt{\left| \sin\frac{\varphi}{2}\right| - \sin^2\frac{\varphi}{2}}\right]^{-2}.
\end{equation}

\section{Conclusions}
We have designed optimal interferometric schemes for implementation of two-qubit linear optical quantum filters operating in the coincidence basis. The considered linear optical realization of the quantum filters may impose an extra cost in terms of reduced success probability 
of successful filtering and the designed schemes maximize the success probability of the filter. The symmetric real filters were analyzed in particular detail  and, interestingly, we have found that for a certain range of parameters the optimal scheme is asymmetric 
in the sense that it couples a pair of modes corresponding to logical state $|1\rangle$ of one qubit and logical state $|0\rangle$ of the other qubit, which contrasts the symmetry of the considered filter. Our investigation of the optimal implementation of optical quantum 
filters complements the earlier studies on optimal realization of linear optical unitary quantum gates. The required interferometric setup can be 
implemented on-chip with integrated optics where a tunable beam splitter can be realized using a Mach-Zehnder interferometer with tunable phase shift~\cite{Carolan2015, Qiang2018, Wang2020}. A universal integrated optics circuit that can realize all of the optimal interferometric schemes 
is drawn in Fig.~\ref{fig-integrated}(a).  As a second example, in Fig.~\ref{fig-integrated}(b) we show a possible bulk optics realization based on polarization qubit encoding and utilization of inherently stable interferometers formed by a sequence of calcite beam displacers~\cite{Lanyon2009, Broome2010, Bian2017, Starek2018B}.
The investigated two-qubit linear optical quantum filters may find applications in linear optics quantum information processing and quantum state engineering.  

\acknowledgments
We acknowledge support by the Czech Science Foundation under Grant No. 19-19189S.

\appendix*
\section{Derivation of structure of  matrix $U_{AB}$}

In this Appendix we determine the most general form of the $4\times 4$ matrix  $U_{j,k}$, $j,k\in{A_0,B_0,A_1,B_1}$,  that describes interferometric coupling which enables 
implementation of the diagonal two-qubit quantum filter (\ref{MfilterLO}).   Recall that input two-photon Fock state $|1_{A_j},1_{B_k}\rangle$ transforms as follows,
\begin{widetext}
\begin{equation}
|1_{A_j},1_{B_k}\rangle \rightarrow \sum_{m,n=0}^1 (U_{A_j,A_m} U_{B_k,B_n}+U_{A_j,B_n}U_{B_k,A_m}) |1_{A_m},1_{B_n}\rangle,
\end{equation}
\end{widetext}
where we assume operation in the coincidence basis and restrict ourselves to the outputs where a single photon is present in each pair of modes $A_0,A_1$ and  $B_0,B_1$. 
Recall also that the implementation of a diagonal two-qubit quantum filter $M=\sum_{j,k=0}^1 m_{jk}|jk\rangle\langle jk|$ requires that 
\begin{equation}
U_{A_j,A_m} U_{B_k,B_n}+U_{A_j,B_n}U_{B_k,A_m}= \sqrt{P_L} m_{jk} \delta_{jm}\delta_{kn}.
\label{UUstructure}
\end{equation}
Throughout the following discussion we assume that all four coefficients $m_{jk}$ are nonzero. Let us first prove that all four diagonal matrix elements $U_{A_j,A_j}$ and $U_{B_k,B_k}$ must be nonzero. Assume that $U_{A_0,A_0}=0$. 
In order to obtain nonzero $m_{A_0,B_0}$ and $m_{A_0,B_1}$ the matrix elements
$U_{B_0,A_0}$, $U_{A_0,B_0}$, $U_{A_0,B_1}$, and $U_{B_1,A_0}$ must be all nonzero. However, this implies that
\begin{equation}
U_{A_0,A_0}U_{B_0,B_1}+U_{A_0,B_1}U_{B_0,A_0} 
\label{Ujjproof}
\end{equation}
is nonzero, which is in contradiction with the required structure (\ref{UUstructure}). Specifically, nonzero  term (\ref{Ujjproof}) implies that the input state $|1_{A_0},1_{B_0}\rangle$
 is transformed to a state that contains non-vanishing contribution of $|1_{A_0},1_{B_1\rangle}$, which is not compatible with the diagonal form of the targeted quantum filter. 
We have thus proved by contradiction that $U_{A_0,A_0}$ must be nonzero. The same proof applies also the the other three matrix elements  $U_{A_1,A_1}$,  $U_{B_0,B_0}$, and $U_{B_1,B_1}$. 

We next show that the four matrix elements $U_{A_0,A_1}$,  $U_{A_1,A_0}$, $U_{B_0,B_1}$, and $U_{B_1,B_0}$ must be zero. We again prove this by contradiction. 
We provide the proof for $U_{A_0,A_1}$. Equation (\ref{UUstructure}) implies that
\begin{eqnarray}
U_{A_0,A_0} U_{B_0,B_1}=-U_{A_0,B_1}U_{B_0,A_0},  \nonumber \\
U_{A_0,A_1} U_{B_0,B_0}=-U_{A_0,B_0}U_{B_0,A_1},  \label{UA0A0proof}\\
U_{A_0,A_1} U_{B_0,B_1}=-U_{A_0,B_1}U_{B_0,A_1} .\nonumber
\end{eqnarray}
If we take the product of the first two equations (\ref{UA0A0proof}) and make use of the third equality (\ref{UA0A0proof}), we obtain 
\begin{equation}
U_{A_0,A_1}U_{B_0,B_1}(U_{A_0,A_0}U_{B_0,B_0}+U_{A_0,B_0}U_{B_0,A_0})=0.
\end{equation}
Since the term in the parentheses is equal to $\sqrt{P_L}m_{00}$ and thus nonzero, we have 
\begin{equation}
U_{A_0,A_1}U_{B_0,B_1}=0.
\label{UA0A0condition1}
\end{equation}
This implies that also 
\begin{equation}
U_{A_0,B_1}U_{B_0,A_1}=0.
\label{UA0A0condition2}
\end{equation}
If $U_{A_0,A_1}\neq 0$, then also $U_{B_0,A_1}\neq 0$, and $U_{A_0,B_1}=0$, which follows from Eqs. (\ref{UA0A0proof}) and (\ref{UA0A0condition1}) and from the above proved condition $U_{B_0,B_0} \neq 0$.
It follows that the amplitude
\begin{equation}
U_{A_0,A_1}U_{B_1,B_1}+U_{A_0,B_1}U_{B_1,A_1}
\end{equation}
is nonzero, although it should vanish. Therefore, $U_{A_0,A_1}=0$ must hold and similarly we can show that also $U_{A_1,A_0}=U_{B_0,B_1}=U_{B_1,B_0}=0$. 

Let us now assume that modes $A_0$ and $B_0$ are interferometrically coupled and $U_{A_0,B_0}\neq 0$, as well as  $U_{B_0,A_0}\neq 0$. We show that the other pairs of modes $A_j$ and $B_k$ 
cannot be coupled.  It follows immediately from Eq. (\ref{UA0A0proof}) that 
\begin{equation}
U_{A_0,B_1}=U_{B_0,A_1}=0.
\end{equation}
We next consider the following amplitudes that should also vanish,
\begin{eqnarray} 
U_{B_1,B_0}U_{A_0,A_0}+U_{A_0,B_0}U_{B_1,A_0}=0,\nonumber \\
U_{A_1,A_0}U_{B_0,B_0}+U_{A_1,B_0}U_{B_0,A_0}=0,
\end{eqnarray}
Since $U_{B_1,B_0}=U_{A_1,A_0}=0$ and $U_{A_0,B_0}\neq 0$,  $U_{B_0,A_0}\neq 0$, we get
\begin{equation}
U_{A_1,B_0}=U_{B_1,A_0}=0.
\end{equation}
Finally, from the requirement that the following two amplitudes should vanish,
\begin{eqnarray}
U_{A_1,B_1}U_{B_0,A_0}+U_{A_1,A_0}U_{B_0,B_1}=0,\nonumber \\
U_{B_1,A_1}U_{A_0,B_0}+U_{A_0,A_1}U_{B_1,B_0}=0,
\end{eqnarray}
we can deduce that 
\begin{equation}
U_{A_1,B_1}=U_{B_1,A_1}=0.
\end{equation}
To summarize our findings: out of the $16$ matrix elements $U_{j,k}$, where $j,k\in\{A_0,A_1,B_0,B_1\}$, only $6$ elements are nonzero: the four diagonal elements $U_{j,j}$
 and two elements representing interferometric coupling of a single pair of modes $A_j$ and $B_k$, e.g. $U_{A_0,B_0}$ and $U_{B_0,A_0}$. The matrix (\ref{UABmatrix00}) considered in Section IIIA of the manuscript 
(or its variants obtained by swapping the modes $A_0$ and $A_1$ and/or $B_0$ and $B_1$) therefore represents the most general permissible interferometric coupling for the implementation of two-qubit diagonal quantum filters.

We note that, strictly speaking, this result holds only if all four $m_{jk}$ are nonzero. If two or three filter parameters $m_{jk}$ vanish, then it can be shown that the filter can be implemented with $P_L=1$ and coupling of one pair of modes is sufficient to achieve this. 
In fact, the only non-trivial configuration is  $m_{01}=m_{10}=0$ while $m_{00}\neq 0$,  and this is covered by the optimal symmetric quantum filters discussed in Section III. Otherwise, $m_{00}=m_{01}m_{10}$ holds, and the filter factorizes into product of two single-qubit filters.
For the remaining case of one vanishing parameter one can conjecture that the dependence of $P_L$ on the filter parameters should be continuous and therefore it should suffice to consider the above identified interferometric configurations with one pair of coupled modes.

\end{document}